\documentclass[prl,aps,showpacs,floatfix,floats,superscriptaddress,nobalancelastpage,twocolumn]{revtex4-1}
\usepackage{graphicx}
\usepackage{graphics}
\usepackage{latexsym,amsmath,amssymb,bm,euscript}
\usepackage{color}
\usepackage{dcolumn}
\usepackage{bm}
\usepackage{epstopdf}
\usepackage{times}
\usepackage{multirow}
\usepackage{wasysym}
\usepackage{subfigure}
\usepackage{bbm}

\def\ra{\rangle}
\def\la{\langle}

\begin{document}

\title{Entanglement Entropy Scaling Laws and Eigenstate Typicality in Free Fermion Systems}
\author{Hsin-Hua Lai}
\affiliation{National High Magnetic Field Laboratory, Florida State University, Tallahassee, Florida 32310, USA}
\author{Kun Yang}
\affiliation{Department of Physics and National High Magnetic Field Laboratory, Florida State University, Tallahassee, Florida 32306, USA}
\date{\today}

\begin{abstract}
We demonstrate that the entanglement entropy area law for free fermion ground states and the corresponding volume law for highly excited states are related by a position-momentum duality, thus of the same origin. For a typical excited state in the thermodynamic limit, we further show that the reduced density matrix of a subsystem approaches thermal density matrix, provided the subsystem's  linear size is small compared to that of the whole system in all directions, a property we dub eigenstate typicality. This provides an explicit example of thermalization via entanglement, and reveals how statistical physics emerges from a single eigenstate by tracing out a large number of degrees of freedom.
\end{abstract}
\maketitle

\textit{Introduction} --Quantum entanglement is one of the most important concepts in the modern physics \cite{QE_RMP}. The most widely used measure of bipartite block entanglement in many-body systems is the entanglement entropy (EE), which is the von Neumann entropy associated with the reduced density matrix (RDM) of a subsystem, obtained by tracing out degrees of freedom outside it. It is generally believed that the EE of ground states of most local Hamiltonians follow the so called ``area law`` \cite{ Eisert_RMP}, which means that when a system is divided into two subsystems, the EE is proportional to the boundary area between these two subsystems. The area law is crucial for the efficiency of density matrix renormalization group and tensor network based variational methods for computing ground state properties. Violations of the area law are rare (other than in quantum critical one dimensional (1D) systems \cite{Calabrese2004}), and also weak in known examples. Above one dimension, the only firmly established examples are free fermion ground states with Fermi surfaces \cite{Wolf,GioevKlich} and coupled harmonic lattice models with Bose surfaces where gapless bosonic excitations live \cite{Lai_EE4EBL}; the violation is logarithmic (i.e., EE is proportional to surface area multiplied by a factor that grows logarithmically with subsystem size) in both cases. Heuristic argument \cite{Swingle} and detailed perturbative calculation \cite{dingprx12} strongly suggest that such a violation also exists in Fermi liquids which takes the same form as that in free Fermi gas, and numerics \cite{Zhang11,kim} suggests similar violations may exist in certain non-Fermi liquid states with Fermi surfaces. Perhaps the strongest violation known thus far is a power-law enhancement of EE in a very special 1D free fermion model involving random long-range hopping \cite{PouranvariYang14}.

Comparatively speaking much less effort has been devoted to studies of EE associated with (highly) excited states (with an {\em extensive} excitation energy that grows linearly with system size) \cite{Manfred2006,Masanes2009}. While it is generally expected that EE should be extensive (i.e. proportional to the volume of the smaller subsystem) in such cases (except in many-body localized states \cite{Nandkishore2014}), explicit examples of this volume law are very rare, and existing results are either of numerical nature or on 1D systems (in fact often both) \cite{Alcaraz2011,Bhattacharya2013,Palmai2014,Vincenzo2009,Ares2014,Molter2014}. Closely related to this is the issue of thermalization\cite{Deutsch_ETH,Srednicki_ETH,Marcos_ETH,Nandkishore2014},
namely the RDM of a (sufficiently) small subsystem approaches a thermal density matrix even when the whole system is in a pure state.
If thermalziation holds then entropic volume law follows, but the opposite is not necessarily true. 
For random initial pure states this is known to be true after a long time evolution in many cases, and termed canonical typicality \cite{Goldstein_typicality,Popescu_typicality,Santos2012,Deutsch2013}. However much less is known if such thermalization occurs if the initial states are exact eigenstates of a local Hamiltonian, which form a very special set in the Hilbert space with zero measure. We dub a term \underline{eigenstate typicality} \cite{Typicality} to characterize such thermalization of an eigenstate, if it occurs. There exist numerical evidence and analytic arguments supporting such eigenstate typicality for a variety of systems. However, the general physical mechanism behind thermalization is unclear, and in particular, it is widely assumed that integrable systems {\it do not} thermalize \cite{Rigol_Srednicki,Deutsch2013,Vedika2014, huse2014}, although there are a few numerical studies suggesting the opposite. \cite{Tatsuhiko2012, Alba2014}

In the present work we address the issues mentioned above, by studying EE and thermalization in the {\it integrable} free fermion systems. We demonstrate that for a ``typical`` highly excited state (in a sense to be specified below), (i) EE follows volume law. (ii) In the limit that the ratio between linear sizes of subsystem and whole system vanishes for {\em all} directions, eigenstate typicality {\it holds} for the subsystem, in sharp contrast to the previous belief. Furthermore, in (i) we show that the area law followed by ground state EE and the volume law for excited state are {\em related} by a position-momentum duality, and thus have the {\em same} origin. The conclusion (ii) is a more striking result, where we find thermalization in the (integrable) free fermion system, in which there are infinitely many conserved quantities (namely occupation number of every momentum state is a good quantum number). Our derivation of (ii) clearly illustrates how statistical physics emerges from a {\em single} eigenstate by tracing out a large number of degrees of freedom. It sheds considerable light on the microscopic origin of thermalization.

\textit{Position-Momentum Duality} --We consider free fermion systems with translational invariance, with Hamiltonian $H = \sum_{j\ell} c^\dagger_j h_{j \ell} c_\ell$, where $c_j(c_j^\dagger)$ is the fermion annihilation (creation) operator at site $j$. For a real-space partition $A$ and its complement $B\equiv \bar{A}$, the RDM $\rho_A$ for any general fermion eigenstate $|F\ra$ takes the Gaussian form \cite{Peschel2003}
\begin{eqnarray}\label{Eq:RDM}
\rho_A = tr_B \left[ | F \ra \la F | \right] = e^{-H_e}, &~~H_e = \sum_{j,\ell} c^\dagger_j \kappa_{e j \ell} c_\ell,
\end{eqnarray}
where the (single-particle) {\em entanglement} Hamiltonian $H_e$ within $A$ is determined {\em exclusively} by the two-point correlation function, $M_{j \ell} \equiv \la F | c^\dagger_j c_\ell | F \ra_A$,	where the subscript $A$ means $j, \ell \in A$, via
\begin{eqnarray}\label{Eq:RDM_h}
\kappa_e = \ln \left[ M^{-1} - \mathbbm{1}_{V_A \times V_A} \right],
\end{eqnarray}
where $\mathbbm{1}$ is a $V_A \times V_A$ identity matrix with $V_A$ being the number of lattice sites inside $A$. Defining $R = \sum_{j\in A} |j\ra \la j|$ as the projection operator onto $A$ \cite{Zhoushen2012,Zhoushen_prb,GioevKlich} and $P = \sum_{{\bf k}\in F} |{\bf k}\ra \la {\bf k}|$ as the projection operator onto the \textit{occupied} states in the momentum space [Brillouin zone (B.Z.)], with $|{\bf k}\ra $ being an momentum eigenstate and also an eigenstate of the original single particle Hamiltonian, we can write
\begin{eqnarray}
M = R P R.
\end{eqnarray}
The position-momentum duality in free fermion systems means that the eigenvalues of $M= RPR$ are exactly identical to the dual matrix $M' \equiv PRP$, as we now demonstrate. For an eigenstate of $M$, $|E_M\ra$, with eigenvalue $\lambda$, $M|E_M \ra = \lambda | E_M\ra$ and $R|E_M \ra = |E_M\ra$ \cite{Qi_duality}, we have
\begin{eqnarray}
M' \left(P|E_M\ra\right) = M' R | E_M \ra = PM |E_M \ra =  \lambda \left( P |E_M\ra \right),~~
\end{eqnarray}
namely $P | E_M \ra$ is an eigenstate of $M'$ with the {\em same} eigenvalue $\lambda$. Denoting the eigenvalue spectrum of $M$ as  spec($M$), we have \cite{Zhoushen2012,Zhoushen_prb,Qi_duality}
\begin{eqnarray}
spec(RPR) = spec(PRP).
\end{eqnarray}
According to Eqs.~(\ref{Eq:RDM})-(\ref{Eq:RDM_h}), the spectrum of RDM, and thus the corresponding EE, can be determined by either $spec(M)$ or $spec(M')$. We take advantage of this duality in the following section.

\begin{figure}[t]
   \centering
   \includegraphics[width=2.5 in]{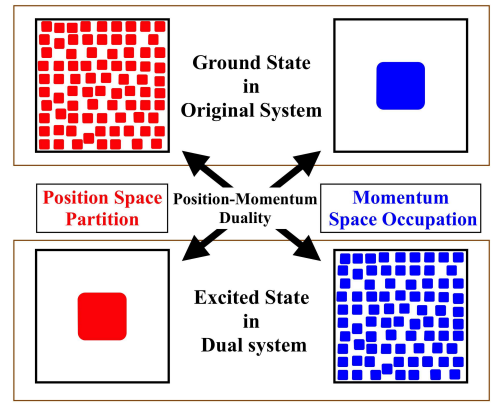}
   \caption{(Color Online) Illustration of position-momentum duality between the original system and its dual, where the roles of momentum and position exchange. In the original system in its ground state (top panel), we consider a fragmented real space partition involving a huge number of pockets distributed over the whole system. The associated Fermi sea of the corresponding ground state is shown in the top right figure. This fragmented partition results in entanglement entropy scaling with the total (sub)system {\it volume}. The dual system (bottom panel) have exactly the same entanglement entropy. However, the momentum occupation corresponds to a highly excited state, while the real-space partition into two contiguous regions is regular. The duality provides a natural understanding of the entropic volume law, expected to be satisfied by typical highly excited states.}
   \label{Fig:duality}
\end{figure}
\textit{Duality between the ground state and the excited state} --
We now show how to relate a ground state to a highly excited state by this duality. Consider a free fermion system in $d$ dimensional Cartesian lattice with total number of lattice sites $V$ in its ground state. The associated Fermi sea in momentum space is shown in the top right panel in Fig.~\ref{Fig:duality}, where the ratio between the number discrete momentum points enclosed in the Fermi sea and that of the whole B.Z. is fixed to be less than but of order 1. We consider a (somewhat unusual) partition in the position space consisting of a huge number of pockets distributed over the whole system that each encloses a large number of lattice sites, top left panel in Fig.~\ref{Fig:duality}. In such a situation the volume ratio between (possibly disconnected) subsystem and total system is held to be a constant when $V$ increases. EE of this special partition can be extracted using known results \cite{GioevKlich}, although for the following discussions we only need to use the area law scaling (with logarithmic correction). If we assume the linear size of each pocket is roughly $L_\square$, EE of this partition can be estimated as $S_P \simeq n L_\square \ln L_\square  \simeq V \ln L_{\square}/L_\square^{d-1} |_{V \rightarrow \infty \gg L_{\square}^d} \sim V$, where we approximate the number of pockets to be $n \simeq V/L_\square^d$. We thus find EE of such a fragmented partition actually scales with the system {\em volume}.

The dual system is illustrated in the lower panel of Fig.~\ref{Fig:duality}, in which real space partitioning and momentum space occupation exchange. We now have in the momentum space a huge number of Fermi pockets distributed in the whole B.Z. This corresponds to a {\em highly excited} state. On the other hand, the position space partitioning is the regular one normally considered in bipartite entanglement.
Using the exact duality discussed above, and the fact that the real and momentum space volumes (as measured by the number of discrete points in them) scale the same way, we conclude that EE of such highly excited states exhibits \textit{volume} instead of area law.

In the above we assumed $L_{\square}\gg 1$ so that we can use the known (area-law) results for ground states. It should be clear, however, that its actually value is unimportant for the volume law to hold. In particular, for a typical highly excited state, we expect $L_{\square}\sim 1$, and volume law should still hold. For a simple illustration, let us consider a highly excited state with {\it staggered} number occupation in the momentum space (even points are occupied and odd points are unoccupied in the B.Z., which is half-filling), in a large 1D chain with total lattice sites $L$. Again EE between a contiguous subsystem $A$ and its complement can be extracted from the matrix $M_{j\ell}\equiv \la c^\dagger_j c_\ell \ra_{ES,A}$, \cite{Peschel2003,Eisler_Peschel} where the subscript $ES$ means we are focusing on an excited state. Once we know the eigenvalues $\lambda_j$ of the matrix $M$, we can obtain EE from $S_{A} = -Tr[\rho_A \ln \rho_A] = - \sum_{j \in A} \left[ \lambda_j \ln \lambda_j + (1-\lambda_j) \ln (1-\lambda_j)\right]$. Regardless of the simplicity of the formula for calculating EE, it is not a trivial task. In most cases, a heavy numerical work needs to be involved. Going around this issue, we instead calculate the particle number fluctuation of $A$: $\la \Delta N^2 \ra_{ES,A}=Tr [ M (\mathbbm{1}-M)]$, which provides the lower bound of the $S_A$. \cite{Klich2006} We consider a subsystem $A$ with fixed $L_A/L = \gamma \leq 1$, in which (we drop the subscript $ES$ to simplify notation)
\begin{eqnarray}
\nonumber && \left\la \Delta N^2 \right\ra_{A} =  Tr \bigg{[} M \left( \mathbbm{1}_{L_A \times L_A} - M \right) \bigg{]}  \\
\nonumber &&=  \sum_{j \in A}  \frac{1}{L} \sum_k n_k  - \sum_{j, \ell\in A} \frac{1}{L^2} \sum_{k, k'} n_k  n_{k'}  e^{-i ( k - k') \cdot (j -\ell)}\\
\nonumber && = \frac{L_A}{L} \sum_k n_k - \frac{1}{L^2} \sum_{ k, k'} n_k n_{k'} \frac{\sin^2\left[ \frac{(k - k')L_A}{2}\right]}{\sin^2\left( \frac{k - k'}{2}\right)}\\
&& = \gamma (1-\gamma)  \sum_m n_m  -  \sum_{m \not = m'} n_m n_{m'}  \frac{\sin^2\left[(m - m')\pi \gamma \right]}{L^2 \sin^2\left[ \frac{(m - m')\pi}{L}\right]},~~\label{Eq:DeltaN}
\end{eqnarray}
where we define $ n_k  \equiv \la c^\dagger_k c_k \ra$, and change the momentum labeling from $k = 2 \pi m /L$ to $m=1,2,\cdots,L$. For the special case of equal partition, we have $\gamma = 1/2$. In this case the second term of Eq.~(\ref{Eq:DeltaN}) {\em vanishes} because of the staggered occupation pattern in momentum space: Since only even momentum points are occupied, we have $n_m n_{m'} = 1$ when $m-m'$ is an even integer {\em only}, resulting in a vanishing numerator for $\gamma=1/2$. We thus find in this case
\begin{eqnarray}
\left\la \Delta N^2 \right\ra_{\rm equal-partition} = \frac{L}{8} \propto L\propto L_A,
\end{eqnarray}
which scales as the subsystem {\it volume}, confirming the heuristic duality picture above. The situation we discussed above is exactly dual to the case studied by Ref.~\cite{Igloi2010} that gives a consistent result to ours.

For an arbitrary excited state with completely {\it random} population in the momentum space, it is not easy to establish a rigorous bound for EE for a generic partition. Instead in the following we will consider appropriate limits in which eigenstate typicality holds, in which case the entropic volume law follows.


\textit{Eigenstate typicality for a typical excited state} --
Entropic volume law is a necessary, but insufficient condition for thermalization, namely the RDM taking form of thermal density matrix corresponding to the original Hamiltonian. In this section we consider the condition under which thermalization occurs for a typical highly excited state. To this end we consider a generic lattice and fermion occupation pattern.
Explicitly the element $M_{j \ell}$ is
\begin{eqnarray}\label{Eq:M_matrix}
M_{j\ell} =\la c^\dagger_j c_\ell \ra_A= \frac{1}{V}\sum_{\bf k} n_{\bf k} e^{-i {\bf k}\cdot \delta {\bf r}_{j\ell\in A}},
\end{eqnarray}
where the occupation number $ n_{\bf k} = \la c^\dagger_{\bf k} c_{\bf k}\ra$ for a typical excites state is $1(0)$ for a occupied (unoccupied) state at momentum ${\bf k}$, with components ${\bf k}_j=2\pi n_j/L_j$, $n_j =1, 2, \cdots, L_j$ and $L_j$ is the linear size along the $j$th direction ($j=1, \cdots, d$). For the moment we set $L_j=L$, corresponding to a (hyper) cubic system.

\begin{figure}[t]
   \centering
   \includegraphics[width=2.5 in]{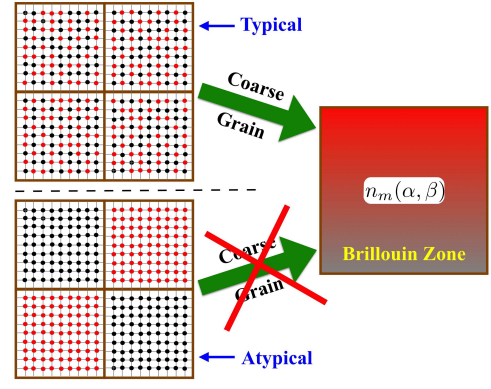}
   \caption{(Color Online) Schematic illustration of the coarse-graining process in momentum space. In the thermodynamic limit, the discrete momentum points become very dense [red (black) dots represent occupied (unoccupied) points], and we can divide the B.Z. into a huge number of cells, the brown boxes, where we only show four such cells for illustration. For a typical excited state (top left panel), the momentum sites are {\it randomly} occupied. Under the coarse-graining process, the {\em average} occupation in each cell varies {\em smoothly} from one cell to the next in the whole B.Z., as illustrated in the right panel; they follow the usual Fermi-Dirac distribution with two parameters $\alpha$ and $\beta$ that are determined by the fixed total energy $E$ and total fermion number $N$. For an atypical excited state in which the cells are not randomly populated (bottom left panel), coarse-graining process will not give rise to a continuous distribution. The probability of encountering such an atypical state vanishes in the limit of large cell size (measured by the number of momentum points it encloses).
}
   \label{Fig:coarse-grain}
\end{figure}

As $L\rightarrow \infty$, the discrete momentum points in the B.Z. become very dense and we can divide the B.Z. into a large number of cells (see left panels in Fig.~\ref{Fig:coarse-grain}).
Each cell contains $g\gg 1$ points associated with the original momenta ${\bf k}$. When $L/L_A \rightarrow \infty$ (we assume the subsystem $A$ is sufficiently isotropic such that it is characterized by a single linear size $L_A$), we can require the linear size of each cell, $\delta {\bf k}_{cell}$ to satisfy
\begin{eqnarray}
1/L \ll \delta {\bf k}_{cell} \ll 1/L_A.
\label{Eq:coarse grain condition}
\end{eqnarray}
With the condition above, for {\it all} the momentum points within the {\em same} cell the phase factor in Eq.~(\ref{Eq:M_matrix}) can be treated as a constant, $\exp[-i {\bf k}_m \cdot \delta {\bf r}_{j \ell \in A}]$, where $m$ is the cell index and ${\bf k}_m$ is its average momentum; we also introduce the corresponding single particle energy $\epsilon_m=\epsilon_{{\bf k}_m}$ for later usage [$\epsilon_{\bf k}$ is band dispersion].
We can thus divide the sum over momenta in Eq.~(\ref{Eq:M_matrix}) into two steps, first summing over momenta within each cell, and then sum over all cells,. We refer to the first step as a ``coarse graining" procedure in momentum space \cite{Kormos2014}, after which the matrix element $M_{j\ell}$ becomes
\begin{eqnarray}
\nonumber M_{j\ell} &\simeq&  \frac{g}{V} \sum_{m} \bigg{(} \frac{N_m}{g} \bigg{)} e^{-i {\bf k}_m \cdot ({\bf r}_j - {\bf r}_\ell )}\\
&=& \frac{1}{V_{eff}} \sum_m n_m e^{-i {\bf k}_m \cdot ({\bf r}_j - {\bf r}_\ell)},
\end{eqnarray}
where $N_m$ is the total occupation number within cell $m$ and $n_m=N_m/g$ is the corresponding average occupation.

It should be clear by now while a specific excited state is characterized by the detailed occupation pattern $\{n_{\bf k}\}$, $M_{j\ell}$ and thus RDM $\rho_A$ depends on the {\em coarse-grained} variables $\{n_m\}$ only. Therefore many {\em different} excited states' will give rise to essentially the {\em same} $\rho_A$, and the {\em most likely} $\rho_A$ corresponds to $\{n_m\}$ consistent with the maximum number of different $\{n_{\bf k}\}$; using standard statistical physics terminology, a specific $\{n_{\bf k}\}$ corresponds to a {\em micro}state, while $\{n_m\}$ corresponds to a {\em macro}state. Based on standard statistical physics arguments, a {\em typical} excited state will result in $\rho_A$ corresponding to this {\em most probable} macrostate $\{n^*_m\}$ in the appropriate limits specified earlier. Let us find out what $\{n^*_m\}$ is.

The only constraints an excited state must satisfy are fixed particle number $N$ and total energy $E$:
\begin{equation}
\begin{array}{lr}
\sum_{m} N_{m} = N, & \sum_{m} N_{m} \epsilon_{m} = E.
\end{array}
\end{equation}
Without these constraints,
for each macrostate $\{ N_m \}$, the number of distinct microstates is denoted as $W\{ N_m\} = \prod_{m} \omega(m)$, where $\omega(m)$ is the number of distinct microstates associated with $m$th cell, $\omega(m) = g !/ [N_m ! (g - N_m)!]$. The number of distinct microstates accessible to the state is $\Omega (N, V, E) = \sum_{\{ N _m \} } W\{N_m \}$, where the summation goes over all the distinct distribution set $\{N_m\}$. The distribution set $\{ N_n \}$ that we are interested is the most probable one and can be obtained by considering the fluctuations of $N_m$ combined with the two constraints above. We introduce Lagrange multipliers $\alpha$ and $\beta$ and examine the fluctuation of the distribution set $\{N_m\}$,
\begin{eqnarray}
\nonumber && \delta \left[\ln W\{N^*_m\} - \left( \alpha \sum_m \delta N_m + \beta \sum_m \epsilon_m \delta N_m \right)\right] = 0\\
&& \Rightarrow n^*_m = \frac{N^*_m}{g}= \frac{1}{e^{\beta\epsilon_m + \alpha} + 1},
\end{eqnarray}
which is exactly the same expression as the Fermi-Dirac distribution from the grand canonical thermal ensemble if we identify $\alpha = -\mu/T$ and $\beta = 1/T$.

The above does not apply, of course, to an atypical excited state like that illustrated in the lower left panel of Fig. \ref{Fig:coarse-grain}. In the limit $g\rightarrow\infty$ which follows from thermodynamic limit $L\rightarrow\infty$, the chance of encountering such states vanishes and we do not consider them further. Using the fact that $n^*$ is a {\em smooth} function in momentum space, the matrix element $M_{j\ell}$ for a {\em typical} excited state approaches
\begin{eqnarray}\label{Eq:M_coarse_grain}
M_{j\ell} = \frac{1}{(2\pi)^d}\int \frac{d^d{\bf k}}{e^{\beta \epsilon_{\bf k} + \alpha} + 1} e^{-i {\bf k} \cdot ( {\bf r}_j - {\bf r}_\ell)}
\end{eqnarray}
for $L/L_A\rightarrow\infty$. We thus find that the RDM of a typical excited state becomes the same as the thermal state density matrix corresponding to the original Hamiltonian, which gives an explicit example of the thermalization \cite{Deutsch_ETH,Srednicki_ETH,Marcos_ETH}.

We remark again that the realization of eigenstate typicality is only valid in the limit that we are considering here, $L/L_A \rightarrow \infty$ (for sufficiently isotropic subsystem), since only in this limit the coarse-graining procedure is well-defined. For highly anisotropic subsystems, we need $L/L_A \rightarrow \infty$ along {\em  all} directions for Eq. (\ref{Eq:coarse grain condition}) to be valid, so that the coarse graining procedures outlined earlier can be followed. This is a slightly more stringent condition than simply having $V/V_A \rightarrow \infty$, which is the normally expected condition for thermalization to hold. We also emphasize that the key step leading to the conclusion above, namely momentum space coarse-graining, is {\em not} an ensemble averaging process; it is averaging the occupation number in a momentum space cell within {\em a single} excited state.  Last, the difference here, as compared to other integrable system, lies in the fact that the conserved quantities (occupation number of every momentum state) {\it do not} have corresponding {\em local} densities.

One diagnostic of thermalization is comparing the von Neumann entropy with the entropy of the thermal state with energy and particle densities corresponding to those of the excited state. If thermalization occurs, these two entropies should be the same, as observed numerically \cite{Storms2014,Stormsnote} under the appropriate conditions specified above. On the other hand, if we fix $V_A/V\sim O(1)$ while taking the thermodynamic limit, thermalization is {\em not} expected to occur. In this case EE, while still following the volume law as demonstrated earlier, does {\em not} approach thermal entropy, as is also seen \cite{Storms2014}. Thus all of our results are fully supported by the numerics of Ref.~\cite{Storms2014}.

\textit{Conclusion}--In this Letter we show that in free fermion systems the entanglement entropy volume law of a typical excited state can be understood from the area law followed by their dual ground states via a position-momentum duality. For the subsystems whose sizes are much smaller than the total system, the reduced density matrix of the subsystem is shown to be the same as in the thermal state via a coarse-graining procedure in momentum space, which we dub eigenstate typiclity. This gives the simplest demonstration of the emergence of the thermalization in free fermion systems.

\textit{Acknowledgments }- We thank Dan Arovas for helpful discussions and Rajiv Singh for a useful correspondence. This research is supported by the National Science Foundation through grants No. DMR-1004545 and No. DMR-1442366.
\bibliography{biblio4EE}
\end{document}